\documentclass{ws-ijmpa}
\def\beq{\begin{equation}}
\def\eeq{\end{equation}}
\def\ba{\begin{array}}
\def\ea{\end{array}}
\def\bea{\begin{eqnarray}}
\def\ena{\end{eqnarray}}

\begin{document}

\markboth{F. Larios, R. Mart\'{\i}nez and M.A. P\'erez}
{New Physics effects in the FCNC of the Top quark.}


%

\catchline{}{}{}{}{}
%


\title{\bf NEW PHYSICS EFFECTS IN THE FLAVOR-CHANGING NEUTRAL
COUPLINGS OF THE TOP QUARK}

\author{F. LARIOS}

\address{
Departamento de F\'{\i}sica Aplicada, CINVESTAV-M\'erida, A.P. 73,\\
97310 M\'erida, Yuc., M\'exico\\
larios@mda.cinvestav.mx}

\author{R. MARTINEZ}

\address{
Departamento de F\'{\i}sica, Universidad Nacional, 
Apartado a\'ereo 14490,\\
Bogot\'a, Colombia\\
remartinezm@unal.edu.co}

\author{M.A. PEREZ}

\address{
Departamento de F\'{\i}sica, Cinvestav, A.P. 14-740, 07000,\\
M\'exico D.F., M\'exico\\
mperez@fis.cinvestav.mx}

\maketitle

\begin{history}
\received{Day Month Year} \revised{Day Month Year}

\end{history}

\begin{abstract}
We survey the flavor-changing neutral couplings (FCNC) of the top
quark predicted by some extensions of the Standard Model: THDM,
SUSY, L-R symmetric, TC2, 331, and models with extra quarks.
Since the expected sensitivity of the LHC and ILC for the tcV
$(V=\gamma,g,Z)$ and tcH couplings is of order of a few percent,
we emphasize the importance of any new physics effect that gives a
prediction for these FCNC couplings within this limit. We also
review the constraints imposed on these couplings from low-energy
precision measurements.
\end{abstract}

\ccode{PACS numbers: 14.65.Ha, 12.15.Mm, 12.60.-i, 12.60.Cn}

\section{Introduction}
The weak neutral current (WNC) was the primary prediction to be
tested in the electroweak $SU(2)\times U(1)$ standard model (SM).
The flavor-conserving structure of the WNC has been verified with
high precision in many processes \cite{pdg}. In the SM, there are no
flavor changing neutral couplings (FCNC) mediated by the $Z$,
$\gamma$, $g$ gauge bosons nor the Higgs boson $H$ at tree level
because the fermions are rotated from gauge to mass eigenstates
by unitary diagonalization matrices \cite{gim}. Furthermore, the
top-quark FCNC induced by radiative effects are also highly
suppressed \cite{rosado,eilamsoni,mele}:
the higher order contributions induced
by the charged currents are proportional to
$(m_i^2 -m_j^2)/M_W^2$, where $m_{i,j}$ are the masses of the quarks
circulating in the loop and $M_W$ is the $W$ gauge boson mass. As
a consequence, in the SM all top-quark FCNC transitions $t\to
qV,qH$ , with $V=Z, \gamma,g$, which involve down-type quarks in
the loops, are suppressed far below the observable level at
existing or upcoming high energy colliders
\cite{rosado,eilamsoni,mele}. For
example, in the $t\to cV$ transitions the scale of the respective
partial widths is set by the $b$ quark mass
\cite{rosado,eilamsoni}, \bea
\Gamma(t\to V_ic) \; =\; |V_{bc}|^2 \alpha \alpha_i \; m_t
\,(\frac{m_b}{M_W})^4 \, (1 - \frac{m^2_{V_i}}{m_t^2}) \ena where
$\alpha_i$ is the respective coupling for each gauge boson $V_i$. 
From the above result, it follows the
approximated branching ratios $BR(t\to \gamma,Z)\sim 10^{-13}$
and $BR(t\to cg)\sim 10^{-11}$. In contrast, in the $b \to s
\gamma$ transitions the leading contribution is proportional to
$m_t^4/M_W^4$ and thus the GIM mechanism \cite{gim} induces in this
case an enhancement factor. In a similar way, it has been
realized that some top-quark FCNC decay modes can be enhanced by
several orders of magnitude in scenarios beyond the SM, and some
of them falling within the LHC's reach. In this case, the
enhancement arises either from a large virtual mass  or from the
couplings involved in the loop. Top-quark FCNC processes may thus
serve as a window for probing effects induced by new physics.

The absence of the vertex Htc at tree-level in the SM can be
traced down to the presence of only one Higgs doublet. The process
involved in the diagonalization of the fermion masses induces
simultaneously diagonal Yukawa couplings for the physical Higgs
boson. In models with more than one Higgs doublet, additional
conditions have to be imposed to ensure that no FCNC arise at
tree level. In particular, a discrete symmetry that makes quarks
of same charge to interact with only one of the two (or more)
Higgs doublets will, by the Glashow-Weinberg mechanism,
cause all the Yukawa couplings involving physical neutral
Higgs boson states become diagonal\cite{glashow}. 
On the other hand,
without any FCNC suppression mechanism these type of models may
produce $tqH$ couplings at tree level, which in turn may
induce large enhancements of the FCNC $tqV$ by radiative
effects \cite{cordero1}.
The interest in FCNC top-quark physics is expected to increase
since processes involving top and Higgs FCNC will be examined
with significant precision at both the LHC and ILC. In the first
case, with a LHC luminosity of $100 fb^{-1}$, 80 million of $t\bar
t$ pairs per year will make it possible to reach the following
limits \cite{aguilar}: 
\bea BR(t\to cH) &<& 6\times 10^{-5}\; ,
\nonumber\\
BR(t\to c\gamma) &<& 1\times 10^{-5}\; ,
\nonumber\\
BR(t\to cZ) &<& 4 \times 10^{-5}\; ,
\nonumber\\
BR(t\to cg) &<& 2\times 10^{-5}\; ,\label{lhcsensitivity}
\ena
while at the ILC, with an
integrated luminosity of $100-200 fb^{-1}$ one can hope to reach
the sensibilities \cite{aguilar,abelc}: \bea
BR(t\to cH) &<& 4.5\times 10^{-5} \nonumber\\
BR(t\to c\gamma) &<& 7.7\times 10^{-6}\; . \ena

The goal of the present review is to bring together much of what
is currently known on top-quark FCNC. We will concentrate on the
predictions made by different extensions of the SM as well as the
constraints imposed on these couplings from low-energy precision
measurements. The physics associated to the production mechanisms
of the processes induced by the top-quark FCNC at future
accelerators have been surveyed in several reviews
\cite{reviews} and will not be considered here.
Our interest is
to compare the predictions on the top-quark FCNC made by the
following models: SUSY, two-Higgs doublets models (THDM),
top-color assisted Technicolor, left-right symmetric models, 331
models, and models with extra quark singlets or extra sequential
quarks. We will introduce first the SM predictions, and then we
will address each one of these models in the following sections.
The last section will be devoted to the analysis of the
model-independent constraints for these couplings.

\section{Top-quark SM couplings}

Although the top quark was discovered ten years ago
\cite{abachi,abecdf},
its couplings to the gauge bosons $\gamma, g, W$ and $Z$ have not
been yet measured directly \cite{reviews}.
Current data provide only
weak indirect limits on the $tbW$ and $ttV$ couplings, with
$V=\gamma, g, Z$. We will use the following parameterization
corresponding to effective $ttV$ interactions with quarks
on-shell and the gauge bosons coupled effectively to massless
fermions \cite{hollik}:
\bea 
\Gamma^{ttV}_{\mu}(k^2,q,\bar q)&=&-ie
\{\gamma_\mu(F^V_{1V}(k^2)+
\gamma_5F^V_{1A}(k^2))+\frac{\sigma_{\mu\nu}}{2m_t}(q+\bar q)^\nu
(iF^V_{2V}(k^2)\nonumber \\
&+&\gamma_5F^V_{2A}(k^2))\}\ena where $m_t$ is the
top quark mass, $q (\bar q)$ is the outgoing top (anti-top) quark
momentum, $k^2=(q+\bar q)^2$, and at tree level in the SM, 
\bea 
F_{1V}^{\gamma,SM}&=&-\frac{2}{3}e   ,  
\;\;\;  \;\;\; \;\;\;  \;\;\; \;\;\;  \;\;\; 
\hspace{3.2cm} F_{1A}^{\gamma,SM}=0, \nonumber \\
F_{1V}^{Z,SM}&=&-\frac{e}{4s_Wc_W}(1-\frac{8}{3}s_W^2)  ,   
 \;\;\; \;\;\; \hspace{2cm}
F_{1A}^{Z,SM}=\frac{e}{4s_Wc_W}, \nonumber \\ 
F_{2V}^{\gamma,SM}&=&F_{2V}^{Z,SM}=0  , 
\hspace{3.5cm}\;\;\; F_{2A}^{\gamma,SM}=F_{2A}^{Z,SM}=0, 
\ena 
where $c_W=\cos\theta_W$,  $s_W=\sin\theta_W$ and $\theta_W$ 
is the weak mixing angle. 
The functions $F_{1V}^V(0)$ and $F_{1A}^V(0)$ are
the $ttV$ vector and axial vector form factors,
$F_{2V}^g(0)=2e/3(g_t-2)/2, F_{2A}^g(0)=2m_td_t^g$, with $g_t$
and $d_t^g$ the magnetic and the (CP-violating) electric dipole
form factors of the top quark. There are similar relations for
$F_{2V}^Z(0)$ and $F_{2A}^Z(0)$ with the weak magnetic and
electric dipole form factors of the Z gauge boson.

It is possible to parameterize possible deviations from the SM
predictions for the $tbW$ and $ttZ$ couplings in terms of only
four coefficients $\kappa_{L,R}^{NC}$ and $\kappa_{L,R}^{CC}$
defined as follows \cite{malkawi}: 
\bea 
{\cal L}&=&\frac{g}{2c_W}\left(1-\frac{4s_W^2}{3}+
\kappa_L^{NC}\right)\bar{t}_L\gamma^\mu t_L Z_\mu 
+\frac{g}{2c_W}\left(\frac{-4s_W^2}{3}+
\kappa_R^{NC}\right)\bar{t}_R\gamma^\mu t_R Z_\mu \nonumber\\
&+&\frac{g}{\sqrt{2}}\left(1+\kappa_L^{CC}\right)
\bar{t}_L\gamma^\mu b_L W^+_\mu 
+\frac{g}{\sqrt{2}}\left(1+\kappa_L^{CC\dagger}\right)
\bar{b}_L\gamma^\mu t_L W^-_\mu \nonumber\\
&+&\frac{g}{\sqrt{2}}\kappa_R^{CC}\bar{t}_R\gamma^\mu b_R W^+_\mu 
+\frac{g}{\sqrt{2}}\kappa_R^{CC\dagger}\bar{b}_R\gamma^\mu t_R W^-_\mu 
\ena 
where $t_L$
denotes a top quark with left-handed chirality, etc. While the
$ttZ$ vector and axial-vector couplings are tightly constrained
by the LEP data \cite{malkawi,lariosyuan},
the right handed $tbW$ coupling is
severely bounded by the observed $b\to s\gamma$ rate 
\cite{perezyuan} at the $2\sigma$ level, 
\bea |Re(\kappa^{CC}_R)| &\leq & 0.4 \times
10^{-2} \nonumber \\
-0.0035& \leq & Re(\kappa^{CC}_R) + 20|\kappa^{CC}_R|^2 \leq
0.0039. 
\ena 
On the other hand, LEP/SLC data also constrains the
other top-quark couplings included in Eq.(4). Even though these
data do not restrict all the anomalous $\kappa$ terms, they
induce the following inequalities 
\bea
-0.019 &\leq & (\kappa^{NC}_R -\kappa^{NC}_L)- 
(\kappa^{NC}_R -\kappa^{NC}_L)^2
+ \kappa^{CC}_L +\kappa^{CC\;\; 2}_L  \leq 0.0013 
\nonumber \\
-0.33 &\leq & (\kappa^{NC}_R -4 \kappa^{NC}_L) 
(1+ 2 \kappa^{CC}_L) \leq 0.1 \nonumber \\
\kappa^{CC}_L &\sim & \kappa^{NC}_L - \kappa^{NC}_R
\ena 
These relations impose in turn strong correlations
on the $\kappa$ couplings so that if only one coupling,
$\kappa_L^{CC}$ for instance, is not zero, the others are forced
to be about the same order of magnitude \cite{perezyuan}.

On the other hand, at an $e^+e^-$ linear collider (ILC) with
${\sqrt s} = 500 \; GeV$, and an integrated luminosity of $100-200
\; fb^{-1}$, it will be possible to measure the $ttV$ couplings in
$t\bar t$ production with a few percent precision \cite{abelc}.
In the LHC, with an integrated luminosity of $30 fb^{-1}$,
it will be possible to probe the $tt\gamma$ coupling with a
precision of $10-35\%$ per experiment \cite{baur}.
The sensitivity limits on the
$ttZ$ couplings will be significantly weaker than those expected
for the $tt\gamma$ couplings. Thus, the ILC will be the best
place to probe the $ttZ$ couplings at the few percent level.

The most general effective Lagrangian describing the FCNC
top-quark interactions with a light quark $q'=u,c$, containing
terms up to dimension five, can be written as \cite{hanhewett} 
\begin{eqnarray}
{\cal L} &=& \bar t \{ \frac{ie}{2m_t} (\kappa_{tq'\gamma} + i\tilde
\kappa_{tq'\gamma} \gamma_5)\sigma_{\mu\nu}F^{\mu\nu} \nonumber \\
&+&\bar t \{ \frac{ig_s}{2m_t} (\kappa_{tq'g} + i\tilde
\kappa_{tq'g} \gamma_5)\sigma_{\mu\nu}\frac{\lambda^a}{2}G^{\mu\nu}_a 
\nonumber \\
&+& \frac{i}{2m_t} (\kappa_{tq'Z} + i\tilde \kappa_{tq'Z}\gamma_5)
\sigma_{\mu\nu}Z^{\mu\nu}  \nonumber \\
&+& \frac{g}{2c_w} \gamma_\mu(v_{tq'Z} + a_{tq'Z}\gamma_5) Z^\mu
\nonumber \\
&+& \frac{g}{2 \sqrt 2} (h_{tq'H} + i {\tilde h}_{tq'H}\gamma_5)H
\}q^\prime.
\end{eqnarray}
where we have assumed also that the top quark and the
neutral bosons are on shell or coupled effectively to massless
fermions. In terms of these coupling constants, the respective
partial widths for FCNC decays are given by \cite{aguilar} 
\bea
\Gamma(t\to qZ)_\gamma&=&\frac{\alpha}{32s_W^2c_W^2}
\left(\mid \kappa_{tqZ}\mid^2 +\mid\tilde\kappa_{tqZ}\mid^2\right) 
\frac{m_t^3}{M_Z^2}
\left[ 1 - \frac{M_Z^2}{m_t^2} \right]^2
\left[ 1 +2 \frac{M_Z^2}{m_t^2}  \right] \nonumber \\
\Gamma(t\to qZ)_{\sigma}&=&\frac{\alpha}{16s_W^2c_W^2} 
\left(\mid v_{tqZ}\mid^2 + \mid a_{tqZ}\mid^2\right) m_t
\left[ 1 - \frac{M_Z^2}{m_t^2} \right]^2
\left[ 2 + \frac{M_Z^2}{m_t^2}  \right]          
\nonumber \\
\Gamma(t\to q\gamma)&=&\frac{\alpha}{2}
\left(\mid\kappa_{tq\gamma}\mid^2 +
\mid\tilde\kappa_{tq\gamma}\mid^2\right) m_t \nonumber \\
\Gamma(t\to qg)&=&\frac{2\alpha_s}{3}
\left(\mid\kappa_{tqg}\mid^2 +\mid\tilde\kappa_{tqg}\mid^2\right) 
m_t    \nonumber \\
\Gamma(t\to qH)&=&\frac{\alpha}{32s_W^2}
\left(\mid h_{tqH}\mid^2 + \mid \tilde h_{tqH} \mid^2\right) 
m_t \left[ 1-\frac{M_H^2}{m_t^2}\right]^2 
\ena

If we use $m_t=178.0 \pm 4.3$ GeV,
$\alpha(m_t)=1/128.921$, $s_W^{2}=0.2342$, $\alpha_S(mt)=0.108$,
$m_H=115 \; GeV$ and the tree level prediction for the leading
$t\to bW$ decay \cite{pdg} 
\bea 
\Gamma(t\to bW)&=&\frac{\alpha}{16s_W^2}
\mid V_{tb}\mid^2 \frac{m_t^3}{M_W^2}
\left[1-3\frac{M_W^4}{m_t^4}+2 \frac{M_W^6}{m_t^6} \right], 
\ena  
an update of the original SM calculations \cite{rosado,eilamsoni,mele}
for the FCNC top-quark branching ratios gives thus
the following results \cite{aguilar} 
\bea
BR(t\to q\gamma)&=&(4.6^{+1.2}_{-1.0}\pm 0.2\pm 0.4^{+1.6}_{-0.5})
\times  10^{-14} \nonumber \\ 
BR(t\to qg)&=&(4.6^{+1.1}_{-0.9} \pm 0.2\pm 0.4^{+2.1}_{-0.7})
\times  10^{-12} \nonumber \\ 
BR(t\to qZ)&\approx&1 \times  10^{-14} \nonumber \\ 
BR(t\to qH)&\approx&3 \times  10^{-15} 
\label{branchings}
\ena where the uncertainties shown in the $t\to
c\gamma,cg$ branching ratios are associated to the top and bottom
quark masses, the CKM matrix elements and the renormalization
scale. These updated results are about one order of magnitude
smaller than the ones previously obtained 
\cite{rosado,eilamsoni,mele}. For the
decays involving the $u$ quark, the respective BR are a factor
$|Vub/Vcb|^{2}\sim 0.0079$ smaller than those shown in (\ref{branchings}).

\section{Two Higgs doublets models}

One of the simplest extensions of the SM adds a new complex
$SU(2)\times U(1)$ scalar doublet to the Higgs sector and it is
known as the Two Higgs Doublet Model (THDM) \cite{glashow}.
There are
three possible versions of this model depending on how the two
doublets couple to the fermion sector. In particular, models I
and II (THDM-I, THDM-II) include natural flavor conservation
\cite{glashow,hunter}, while model III (THDM-III) has the simplest
extended Higgs sector that naturally introduces FCNC at the
tree level \cite{sher,antara,reina}.
The most general THDM scalar potential, which is
invariant under both SU(2)xU(1) and CP symmetries, is given by
\cite{hunter} 
\begin{eqnarray}
 V(\phi_{1},\phi_{2})& & =  \lambda_{1} ( |\phi_{1}|^2-v_{1}^2)^2
+\lambda_{2} (|\phi_{2}|^2-v_{2}^2)^2+
\lambda_{3}((|\phi_{1}|^2-v_{1}^2)+(|\phi_{2}|^2-v_{2}^2))^2 
+\nonumber\\ [0.2cm]
& &\lambda_{4}(|\phi_{1}|^2 |\phi_{2}|^2 - |\phi_{1}^+\phi_{2}|^2  )+
\lambda_{5} [\Re e(\phi^+_{1}\phi_{2})
-v_{1}v_{2}]^2+ \lambda_{6} [\Im m(\phi^+_{1}\phi_{2})]^2\, . 
\end{eqnarray}
Where $\phi_1$ and $\phi_2$ are the two
Higgs doublets with weak hypercharge Y=1, $v_1$ and $v_2$ are
their respective vacuum expectation values, and the six $\lambda$
parameters are real. The dimension-two term
$\lambda_5Re(\phi_1\phi_2)$ violates softly the discrete symmetry
$\phi_i\to -\phi_i,$ which is essential to induce natural flavor
conservation in models I and II \cite{glashow}.

After the electroweak symmetry breakdown, three of the original
eight degrees of freedom associated to $\phi_1$ and $\phi_2$
correspond to the three Goldstone bosons $(G^{\pm}, G^o)$, while
the other five degrees of freedom reduce to five physical Higgs
bosons: h, H (both CP-even), A (CP-odd), and $H^\pm$. The
combination $v^2= v_1^2+v_2^2$ is fixed by the electroweak scale
$v=(\sqrt{2}G_F)^{-1/2}$ and there are still 7 independent parameters,
which are given in terms of four physical scalar masses $(m_h,
m_H, m_A, m_{H^\pm})$, two mixing angles ($\tan \beta = v_1/v_2$
and $\alpha$) and the soft breaking term $\lambda_5$.

\begin{figure}[pb]
\centerline{\psfig{file=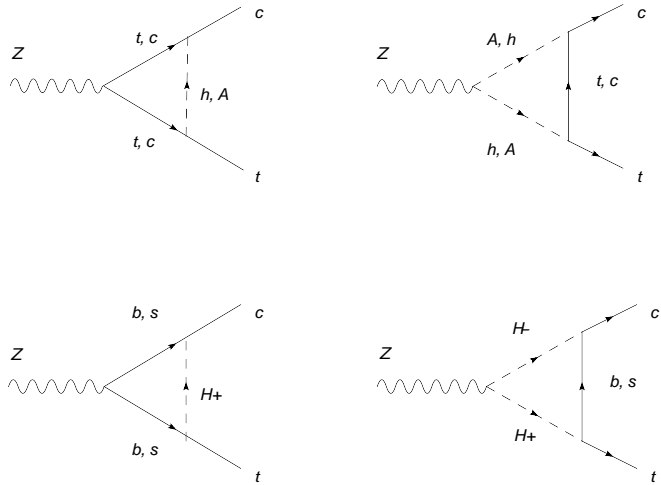,width=8.7cm}}
\caption{Generic contribution to the $tcZ$ and $tc\gamma$
vertices in the THDM-III.}
\label{figdiagramsthdm}
\end{figure}

The three versions of the THDM are distinguished by their Yukawa
couplings, 
\begin{eqnarray}
{\cal L}^{THDM}&=&\eta _{ij}^{U}
\overline{Q}_{iL}\widetilde{\phi }_{1}U_{jR}+
\eta _{ij}^{D}\overline{Q}_{iL}\phi _{1}D_{jR} \nonumber \\ 
&+&\xi _{ij}^{U}\overline{Q}_{iL}\widetilde{\phi }_{2}U_{jR}+\xi
_{ij}^{D}\overline{Q}_{iL}\phi _{2}D_{jR} + h.c. 
\end{eqnarray}
where $\xi^{UD}_{ij}=0$ for model I and
$\eta ^U_{ij}=\xi^D_{ij}=0$ for model II. For model III all
$\eta$'s and $\xi$'s are different from zero. While models I and II
do not generate FCNC at tree level, due to the Glashow-Weinberg
theorem \cite{glashow},
in model III the following FCNC interaction is
obtained after the spontaneous symmetry braking
$<\phi_1>=(0,v/\sqrt{2})$, $<\phi_2>=0$, 
\bea 
{\cal L}^{THDM}=\xi_{ij}\sin\alpha 
\bar f_if_jh+\xi_{ij}\cos\alpha \bar f_if_jH+
\xi_{ij}\cos\alpha \bar f_i\gamma_5f_jA + h.c.
\ena 
In THDM-I and THDM-II the FCNC decay modes $t\to cV$ are dominated
by the one-loop diagrams with a virtual $H^\pm$ and their
respective branching ratios are only sensitive to $m_{H^\pm}$ and
$tan\beta$. The largest enhancements are found for THDM-II with
$120\; GeV < m_{H^\pm} < 250\; GeV$ and $tan\beta>10: BR(t\to
c\gamma)\sim 10^{-7}-10^{-10}$, $BR(t\to cg)\sim 10^{-5}-10^{-9},
BR(t\to cZ) \sim 10^{-8}-10^{-11}$~\cite{eilamsoni,arhrib}.

In the case of the FCNC decay mode with a light Higgs scalar
$t\to ch$, the enhancement is spectacular $BR(t\to ch)\sim 8\times
10^{-5}$ for large $\tan\beta$ and a light charged Higgs mass.
This time the pure scalar couplings $hH^+H^-$ and $hG^+H^-$ play a
crucial role~\cite{guasch03}.

On the other hand, in the THDM-III the heavy neutral scalar and
pseudoscalar Higgs bosons H and A have non-diagonal couplings to
fermions at the tree level. The FCNC decay modes $t\to cV$ and
$t\to ch$ proceed at the one-loop level due to the exchange of $H,
A$ and $H^{\pm}$ (Fig.~\ref{figdiagramsthdm}).
The respective branching ratios may be
enhanced by several orders of magnitude, for reasonable values of
the THDM-III parameters, with respect to the SM predictions:
$BR(t\to cg)\sim 10^{-4}-10^{-8}, BR(t\to c\gamma)\sim
10^{-7}-10^{-11}, BR(t\to cZ)\sim 10^{-6}-10^{-8}$
\cite{reina,iltan,toscano99}.

\section {Supersymmetric models}

Theories with low-energy SUSY have emerged as the most attractive
candidates for physics beyond the SM~\cite{chung}. In particular,
they provide an elegant resolution of the hierarchy problem: SUSY
relates the scalar and fermionic sectors. Thus, the chiral
symmetries which protect the masses of the fermions also protect
the scalar masses from quadratic divergences. In the unbroken
SUSY world, each known particle has a superpartner that differs
in spin by 1/2 and is related to the original particle by a SUSY
transformation. However, SUSY must be a broken symmetry. Otherwise,
the masses of all new superpartners would be equal to the known
particle spectrum. Therefore, the effective Lagrangian at the
electroweak scale is expected to be parameterized by a general set
of SUSY-breaking terms if the attractive features of SUSY are to
be a part of the physics beyond the SM. This version of SUSY is
known as the minimal Supersymmetric standard model (MSSM).

The source of flavor violation in the MSSM arises from the
possible misalignment between the rotations that diagonalize the
quark and squark sectors. The superpotential of the MSSM
Lagrangian is given by

\beq W = \lambda^L_{ij}L_iE^c_jH_1 + \lambda^D_{ij}H_1Q_iD^c_j +
\lambda^U_{ij} U^c_iQ_jH_2 + \mu H_1H_2 \eeq where $L_i$ and
$Q_i$, $i = 1,2,3$ are the chiral superfields for the $SU(2)_L$
doublets for leptons and quarks, $E^c, D^c$ and $U^c$ correspond
to the respective $SU(2)_L$ fermion singlets. $H_1$ and $H_2$
represent two $SU(2)$ Higgs doublets with hypercharge $-1$ and
$+1$, respectively. The MSSM with explicit R-parity violation
includes the following terms in the superpotential:

\beq W_R = \lambda_{ijk}L_iL_jF^c_k + \lambda^{'}_{ijk}L_iQ_jD^c_k
+\lambda^{''}_{ijk}U^c_iD^E_jD^c_k\, .\eeq 
These terms generate a direct
violation of R-parity invariance $(-1)^{3B+L+2S}$ with $B$ and
$L$ the leptonic and baryonic quantum numbers, and $S$ the spin
of each field. The known bounds on the proton decay and the
low-energy FCNC precision measurements set strong constraints on
the $\lambda_{ijk}$ couplings \cite{chung}. The soft-SUSY-breaking
terms responsible for the non-minimal squark family $(\tilde Q,
\tilde n, \tilde d)$ mixing are given by
\bea {\cal L}_{soft} &=& (m^2_{\tilde Q})_{ij} 
\tilde Q^i\tilde Q^j + (m^2_D)_{ij}\tilde U^{i+} \tilde U^j
+ (\tilde m^2_{\tilde D})_{ij} \tilde D^{i+} \tilde D^j \nonumber \\ 
&+& A^U_{ij} \tilde Q^i\tilde U^j_R h_2 + A^D_{ij} 
\tilde Q^i\tilde D^j_R h_1 .
\ena

The FCNC effects come from the non-diagonal entries in the
bilinear terms $M^2_{\tilde Q}, M^2_Q$ and $M^2_{\tilde Q}$, as
well as from the trilinear terms $A^U$ and $A^D$. If this model
becomes universal in the three families at the GUT scale, then we
have
\beq m^2_Q = m^2_{\tilde U} = m^2_{\tilde D} \equiv m^2_o 
\nonumber
\eeq
and
\beq A^{U,D}_{ij} = A_o \lambda^{U,D}_{ij}. 
\eeq 

As far as the top-quark FCNC are concerned, the decays $t\to cV$
have been studied extensively in the MSSM.
The first studies~\cite{oakes} considered one-loop SUSY-QCD and
SUSY-EW contributions,
which were later generalized in order to include the left-handed
(LH) and right-handed (RH) squarks mixings \cite{couture}.
The SUSY-EW
corrections were further generalized and included the neutralino-
$q\tilde q$ loop~\cite{nanopoulos},
as well as the relevant SUSY mixing
angles and diagrams involving a helicity flip in the gluino
line~\cite{petronzio,guasch99}.
While the first calculations obtained $BR(t\to cV)$
of the order of $10^{-6}-10^{-8}$, every new study
improved these results until the range of values
$BR(t\to cg)\sim 10^{-5}$,
$BR(t\to c\gamma)\sim 10^{-6}$, $BR(t\to cZ) \sim 10^{-6}$
were reached.
However, they are still below the estimated sensitivity at the
LHC with an integrated luminosity of 100 $fb^{-1}$
(see Eq.~\ref{lhcsensitivity}).
Similar results were
obtained in a MSSM with a light right-handed top-squark and a
large mixing between the first or second and the third generation
of up-squarks~\cite{delepine}.

Recently, the FCNC top-quark decays have been re-analyzed in the
so-called unconstrained MSSM~\cite{misiak},
where the assumptions on
the soft breaking terms are relaxed and new sources of flavor
violation appear in the sfermions mass matrices. In this case the
neutralino-$q\tilde q$ and gluino-$q\tilde q$ couplings induce
larger contributions to the FCNC processes:
$BR(t\to c\gamma)\sim 10^{-6}$,
$BR(t\to cZ)\sim 10^{-6}$, $BR(t\to cg)\sim 10^{-4}$, with the
last one probably measurable at the LHC~\cite{liuli}.  
If these top-quark decays are not observed at the LHC, upper
bounds will be set on specific soft SUSY breaking parameters.

Another enhancement has been reported~\cite{young} for the
$t\to cV$ decays induced by B-violating couplings in broken
R-parity MSSM: $BR(t\to cg)\sim 10^{-3}$,
$BR(t\to c\gamma)\sim 10^{-5}$,
$BR(t\to cZ)\sim 10^{-4}$, which are definitely within the
LHC's reach (Eq.~\ref{lhcsensitivity}).

On the other hand, while the $t\to cH$ decay is the less favored
channel in the SM~\cite{eilamsoni,mele},
it is this FCNC channel which shows
the most dramatic enhancements due to new physics effects. In
some SUSY extensions its BR can be ten orders of magnitude larger
than the SM prediction. This possibility arises not only because
in some models the FCNC vertex tcH can be generated at tree
level, but also because the GIM suppression does not apply in some
loops. In particular, the gluino-mediated FCNC couplings
$u_a\tilde u_b\tilde g$ induces a $BR(t\to ch)\sim 10^{-4}$,
where $h$ is the lightest CP-even Higgs boson predicted in the
MSSM~\cite{petronzio,guasch99,yangli}.
The branching fraction for this channel has been
also found to be as large as $10^{-3}-10^{-5}$ in a minimal SUSY
FCNC scenario in which all the observable FCNC effects come from
squark mixings $\tilde c - \tilde t$ induced by the non-diagonal
scalar trilinear interactions~\cite{cruzyuan}.  However, it has
been pointed out recently that the electroweak precision
measurements may impose constraints on this squark mixing; which
in turn decrease the MSSM prediction for the FCNC top quark
processes~\cite{cao}.  

If R-parity violation is included in the MSSM , $t\to ch$ receives
new contributions from the loops with an exchange of a single
sparticle that involves the third generation of fermions. As a
consequence, the mass suppression is less severe than in the
purely MSSM~\cite{eilam01}.
In this case the respective branching ratio
can be as high as $10^{-5}$ in some part of the parameter space.
It should be mentioned that in all these cases, the $BR(t\to ch)$
falls off quickly for heavier sparticles in the loops.

The occurrence of tree-level FCNC has also been studied in
supersymmetric multi-Higgs doublet models~\cite{escudero}.
As expected,
it was found that the matrices which diagonalize the quark mass
matrices do not, in general, diagonalize also the corresponding
Yukawa couplings. As a consequence, both scalar and pseudoscalar
Higgs-quark-quark interactions may exhibit a strong
non-diagonality in flavor space. For example, one-loop
contributions to low-energy observables require a Higgs spectrum
at least of order 10 TeV in order to suppress appreciably FCNC of
light quarks~\cite{escudero}.

\section {Topcolor-assisted Technicolor}

In Technicolor theories, the electroweak symmetry breaking
mechanism arises from a new, strongly coupled gauge interaction
at TeV energy scales~\cite{susskind}.
Quark and lepton mass matrices
appear from the embedding of Technicolor in a larger gauge
theory, extended Technicolor~\cite{dimopoulos},
which must be broken
sequentially from energies of order $10^3$ TeV down to the 1 TeV
level. However, the simplest QCD-like extended Technicolor (ETC)
model leads into problems with the LEP precision measurements
data~\cite{pdg}. The topcolor scenario was proposed in order to make
the predictions consistent with the LEP data and to explain the
large top quark mass. Topcolor-assisted Technicolor (TC2) models
\cite{lane}, flavor-universal TC2 models~\cite{popovic},
top see-saw models~\cite{dobrescu},
and the top flavor see-saw models~\cite{tait} are
examples of the topcolor scenario~\cite{simmons}.

In the TC2 model, the topcolor interactions give rise to the main
part of the top quark mass $(1-\epsilon)m_t$, with the model-
dependent parameter $\epsilon$ in the range
$0.03<\epsilon<0.1$~\cite{lane}.
The ETC interactions are responsible for the remaining
part of the top quark mass, $\epsilon m_t$. This model predicts
three heavy top-pions $(\pi_t^{0}, \pi_t^{\pm}$ and one
top-Higgs boson $h^0_{t}$ with large Yukawa couplings to the third
generation of fermions, which thus can induce new top quark FCNC.
Early calculations, in the framework of the simplest ETC models,
already produced large enhancements for the FCNC decay modes
$t\to cV$~\cite{wang}. In the TC2 model, there were also found
\cite{lu} large enhancements for these decay modes which arose
from the virtual contributions of the top-pions and the top-Higgs
boson for reasonable values of the TC2 parameters:
$BR(t\to cg)\sim 10^{-5}$, $BR(t\to cZ)\sim 10^{-5}$,
$BR(t\to c\gamma)\sim 10^{-7}$.

The contribution of an extra neutral gauge boson $Z'$ to the
$t\to c\gamma$ decay mode has been studied also in the framework
of the TC2 model and the so-called 331 model~\cite{pisano}.
Even though the $Z'$ boson predicted in these models couples in
a non-universal way to the third generation of fermions, it was
found that its contribution to the branching ratio of
$t\to c\gamma$ is at most of order of $10^{-8}$ for
$m_{Z'}\sim 500$ GeV~\cite{yueliu}.

\section {Left-Right Symmetric Models}

Left-Right (LR) symmetric models are based on the gauge group
$SU(2)_L\times SU(2)_R\times U(1)_{B-L}$. Their general aim is to
understand the origin of parity violation in low-energy weak
interactions. This gauge symmetry allows a seesaw mechanism and
predicts naturally neutrino masses and mixing~\cite{mohapatra}.
FCNC top-quark decays have been studied in the alternative LR
symmetric model~\cite{davidson},
which is a new formulation of these
models with an enlarged fermion sector: it includes vector-like
heavy fermions in order to explain the fermionic mass hierarchy.
Because of the presence of extra quarks, the CKM mass matrix is
not unitary and FCNC may exists at tree level. In particular,
there is a top-charm mixing angle which induces the tree level
couplings tcZ and tcH. Precision measurements at LEP impose
rather weak constraints on this mixing angle, which in turn allows
FCNC branching ratios as high as
$BR(t\to cH)\sim 10^{-4}$~\cite{gaitan}.

The $t\to cV$ decay modes have been analyzed in two SUSY versions
of LR symmetric models: the constrained or flavor-diagonal case,
in which the only source of flavor violation comes also from the
CKM mass matrix in the quark sector, and the unconstrained
model, in which soft SUSY breaking parameters are allowed to
induce flavor-dependent mixings in the squark mass
matrix~\cite{frankturan}.
The respective branching ratios were calculated in both
cases at the one-loop level with contributions arising from
virtual squarks, gluinos, charginos and neutralinos. In the
flavor- diagonal case, the FCNC top quark branching ratios can
not exceed $10^{-5} (10^{-6})$ for the gluon $(\gamma/Z)$ decay
mode. On the other hand, in the unconstrained LR SUSY model, where
flavor-changing elements in the squark mass matrix are allowed to
be arbitrarily large only for the mixing between the second and
third generations, there are more favorable enhancements:
$BR(t\to cg)\sim 10^{-4}, BR(t\to cZ)\sim 10^{-5}$ and $BR(t\to
c\gamma)\sim 10^{-6}$~\cite{frankturan}.

\section {Models with Extra Quarks}

In models with extra quarks, the CKM matrix is no longer unitary
and the $tcZ$ and $tcH$ couplings may arise at the tree level.
When the new quarks are $SU(2)_L$ $Q=2/3$ singlets, present
experimental data allow large branching ratios: $BR(t\to cZ)\sim
1.1\times 10^{-4}$ and
$BR(t\to cH)\sim 4.1\times 10^{-5}$~\cite{aguilar}.
The decay rates for $t\to cg,c\gamma$ are induced at
the one-loop level but they receive only moderate enhancements:
$BR(t\to cg)\sim 1.5\times 10^{-7}$ and $BR(t\to c\gamma)\sim
7.5\times 10^{-9}$. In models with $Q=-1/3$ quark singlets, the
respective branching ratios are much smaller since the breaking
of the CKM unitarity is very constrained by experimental
data~\cite{aguilar}.
The contributions arising from a sequential fourth
generation $b'$ to the FCNC top-quark decays have been also
studied~\cite{rosado,hou}.
However, the virtual effects induced by a $b'$ heavy quark cannot
enhance the respective branching ratios to within the LHC's
reach: $BR(t\to cZ)\sim 10^{-6}$, 
$BR(t\to cH)\sim 10^{-7} - 10^{-6}$, $BR(t\to cg)\sim 10^{-7}$,
$BR(t\to c\gamma)\sim 10^{-8}$~\cite{hou}.

\section {Three-body decays}

\begin{figure}[pb]
\centerline{\psfig{file=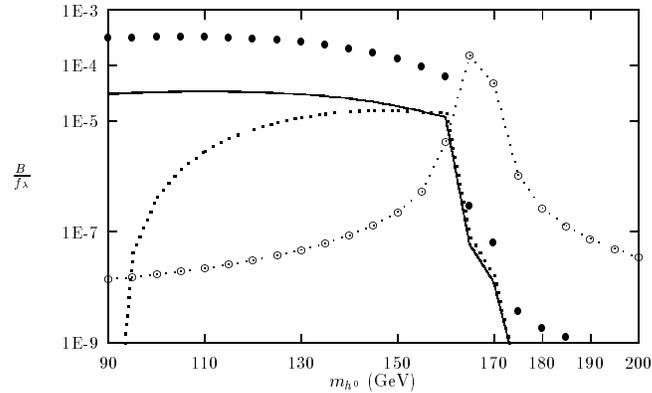,width=8.7cm}}
\caption{Scaled branching ratios for $t\to cV_iV_j$ in the
THDM-III: $t\to c\gamma\gamma$ (solid line), $t\to c\gamma Z$
(points), $t\to cWW$ (hollow circles), and $t\to cgg$ (full
circles). The masses of the Higgs bosons $H, A$ and $H^{\pm}$
were set to $750 \; GeV$.}
\label{figbranchthdm}
\end{figure}

The interest in FCNC three-body decays of the top quark relies in
the phenomenon known as  ``higher order dominance", observed in
b-physics, in which a higher order dominates over a lower order
rate. Of course, the enhancements obtained in two-body FCNC
decays might also appear in these decay modes in some extensions
of the SM~\cite{cruzlopez}.
According to the recent CDF/DO analysis based
on the Tevatron RUN II data~\cite{cdfd0},
the following three-body
rare decays of the top quark may be allowed kinematically: $t\to
bWZ$, $t\to cWW$, $t\to cViVj$ $(Vi=g,Z,\gamma)$, $t\to
c\ell_i\ell_j$, $t\to cu_i\bar u_j$. In particular, the decay
$t\to cZZ$ can only occur through finite-width effects in some
range of the allowed parameter space of THDM-III~\cite{shalom}.

In the SM, only the two decay modes $t\to bWZ$ and $t\to cWW$
arise at the tree-level with branching ratios of order
$10^{-12}-10^{-14}$~\cite{jenkins}.
Since these decay channels ocurr
near the kinematical threshold, the finite decay width of the W
and Z bosons induce sizeable enhancements on the respective
branching ratios: $BR(t\to bWZ)\sim2\times 10^{-6}$ in the SM
\cite{altarelli},
and $BR(t\to cWW)\sim10^{-3}-10^{-4}$ and $BR(t\to
cZZ)\sim 10^{-3}$ in the THDM-III~\cite{shalom}.

The decay modes $t\to cV_iV_j$ have been estimated in the SM
assuming that the respective decay rates are dominated by a
Higgs-boson resonant diagram with a FCNC tcH vertex and the
further two-body decay of the Higgs boson
$H\to V_iV_j$~\cite{toscano99}.
Using the SM values for the effective vertices tcH and
$HV_iV_j$~\cite{eilamsoni,mele,hunter},
one gets the expected SM suppressed values:
$BR(t\to c\gamma\gamma)$, $BR(t \to c\gamma Z)\sim10^{-15}-
10^{-16}$ and
$BR(t\to cgg)\sim10^{-14}-10^{-15}$~\cite{toscano99}.
In the latter case, a complete one-loop calculation in the SM gives
$BR(t\to cgg)=1.02\times 10^{-9}$, two orders of magnitude higher
than the two-body decay rate
$BR(t\to cg)=5.73\times 10^{-12}$~\cite{eft1}.

These decay modes have been studied also in the
framework of the THDM-III within the same Higgs-boson resonant
exchange approximation~\cite{toscano99,shalom97}.
In this case the enhancements
obtained are sizeable due to the combined effect of the
tree-level tcH coupling and the resonance of the intermediate
Higgs boson: $BR(t\to c\gamma\gamma)$, $BR(t\to c\gamma Z) \sim
10^{-4}-10^{-5}$ and
$BR(t\to cWW), Br(t\to cgg) \sim 10^{-4}$~\cite{toscano99}
(Fig.~\ref{figbranchthdm}).
In the latter case, a complete one-loop
calculation in the MSSM produces $BR(t\to cgg)\sim
10^{-7}-10^{-9}$, for reasonable values of the MSSM parameters,
and $BR(t\to cgg), BR(t\to cg)\sim 10^{-5}$ if the SUSY FCNC
couplings are allowed to be large~\cite{eft2}.

Models with additional Higgs triplets can include a tree-level
vertex HWZ, the decay mode $t\to bWZ$ may then proceed by an
intermediate charged Higgs boson, and the same resonance effect
may induce a spectacular enhancement
$BR(t\to bWZ)\sim 10^{-2}$~\cite{cruzlopez}.
A similar situation happens with the radiative three
body decay $t\to ch\gamma$ in the THDM-III, which also can
proceed at the tree level with a large enhancement $BR(t\to
ch\gamma)\sim 10^{-5}$ with respect to the SM prediction $BR(t\to
ch\gamma)\sim 10^{-15}$~\cite{cordero2}.

In TC2 models, the $W^+W^-$ decay channel gets a substantial
enhancement: $BR(t\to cWW)\sim 10^{-3}$, but in the other
$V_i V_j$ decay modes the enhancement is very small~\cite{yue01}.
Finally, the three-body FCNC
decay modes involving a lepton or a quark pair have been
calculated in the THDM-III and TC2 models, but the respective
branching ratios are out of the LHC's reach:
$BR(t\to cq\bar q)$,
$BR(t\to c\ell_i\ell_j) \sim 10^{-7}-10^{-10}$~\cite{eft1,yuewang}.

\section {Constraints from loop observables}

\begin{figure}[pb]
\centerline{\psfig{file=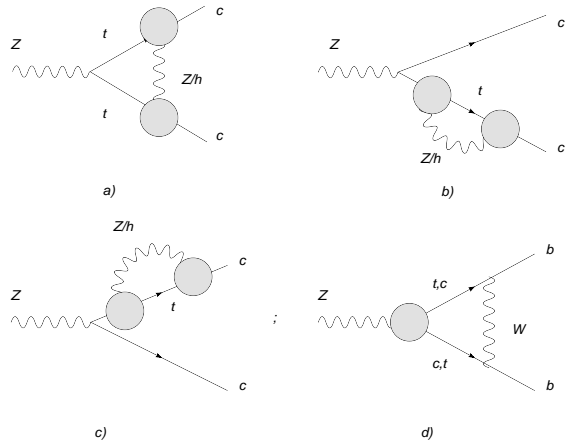,width=7.7cm}}
\caption{Feynman diagrams for the one-loop contribution of the
$tcZ/tcH$ vertices to the decay modes $Z\to b\bar b_,c\bar c$.}
\label{figdiagramszcc}
\end{figure}

\begin{figure}[pb]
\centerline{\psfig{file=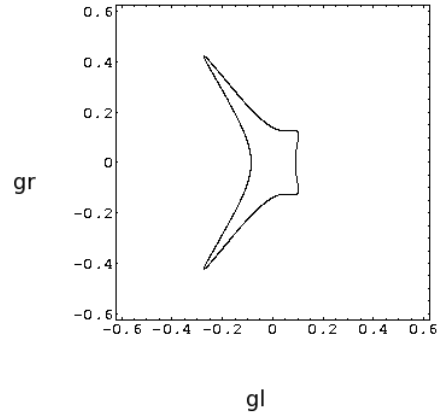,width=7.7cm}}
\caption{A 95\% C.L. fit on the bounds of the dimension-4 $tcZ$
couplings obtained from the current values of the electroweak
precision observables.}
\label{figboundstcz}
\end{figure}

In the effective Lagrangian approach, the new physics effects
induced by non-standard particles can be parameterized as coupling
constants of effective operators which are constructed out of SM
fields~\cite{cruzperez}.
The SM Lagrangian is modified by the addition of
a series of SM-gauge invariant operators with coefficients
suppressed by inverse powers of $\Lambda$, the lowest new-physics
scale. The largest contribution to top-quark FCNC comes from
dimension-6 operators since dimension-5 operators violate lepton
number~\cite{buchmuller}.
After the spontaneous symmetry breakdown, the
dimension-6 operators induce the most general effective
Lagrangian given in Eq. (9), which describes the FCNC top-quark
interactions with a light quark c or u and the gauge bosons
$V=\gamma,g,Z$ and the SM Higgs boson H.

\begin{figure}[pb]
\centerline{\psfig{file=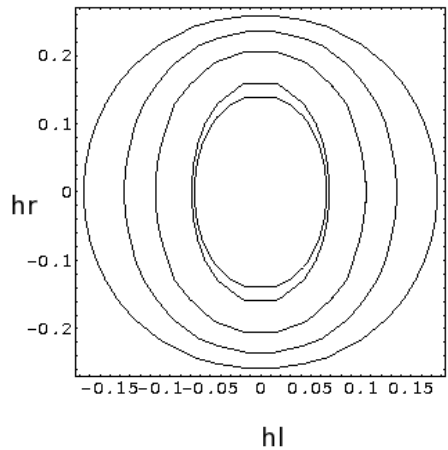,width=6.7cm}}
\caption{A 95\% C.L. fit on the bounds of the dimension-4 $tcH$
couplings obtained from the current values of the electroweak
precision observables. From the inner to the outer curves, the
corresponding values for the Higgs boson mass are:
114, 130, 145, 150 and 160 GeV.}
\label{figboundstch}
\end{figure}

\begin{figure}[pb]
\centerline{\psfig{file=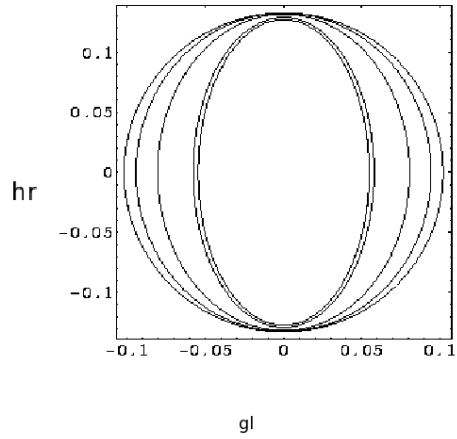,width=9.9cm}}
\caption{A 95\% C.L. fit on the dimension four couplings $h_r$ and
$g_l$ obtained from the interference of the contributions of the
$tcH/tcZ$ effective vertices to the electroweak LEP precision
observables. We fixed $g_r=h_l=0.05$ and the values used for 
the Higgs mass are indicated in
figure~\ref{figboundstch}.} \label{figboundstchz}
\end{figure}

\begin{figure}[pb]
\centerline{\psfig{file=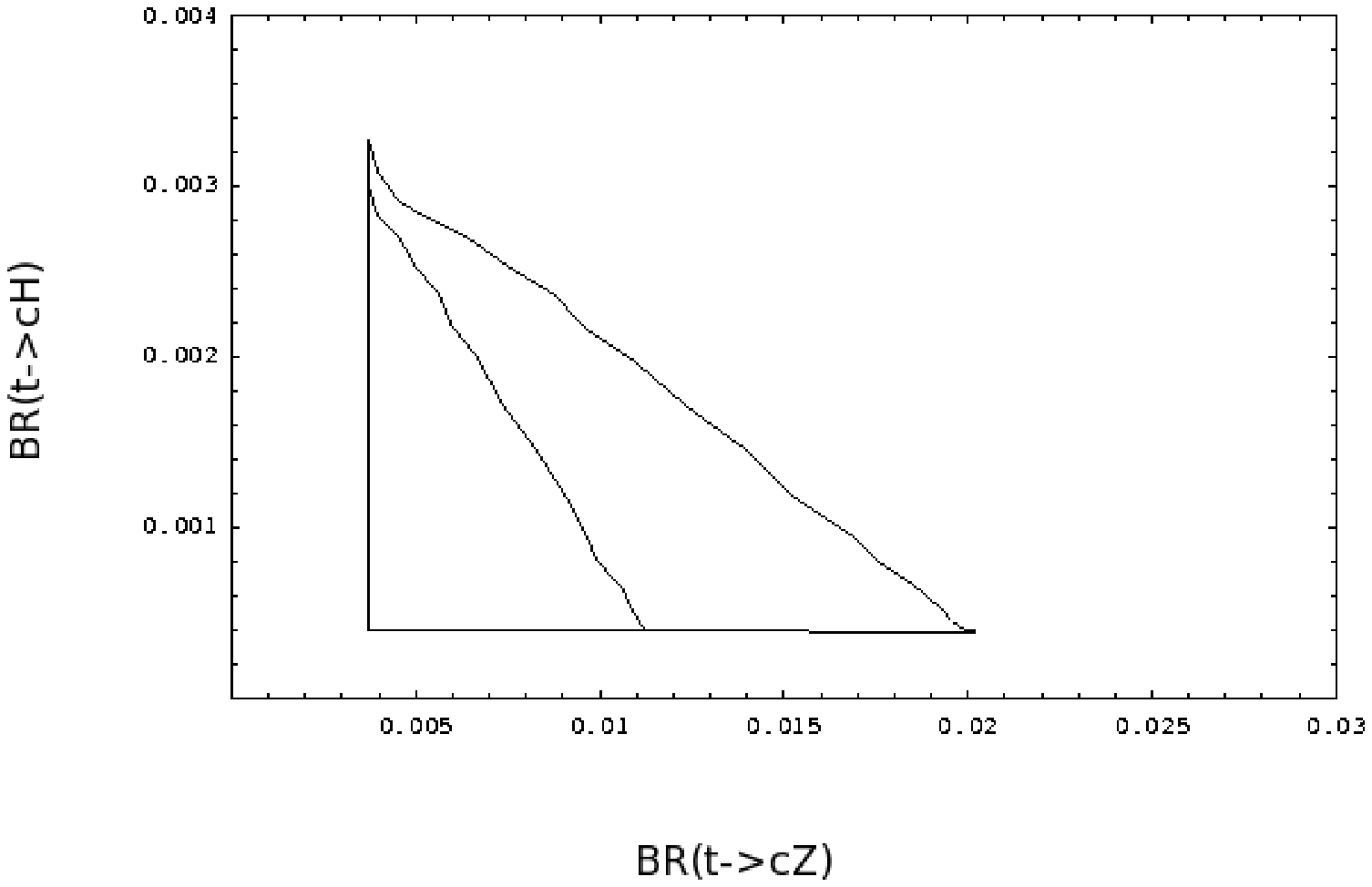,width=9.7cm}}
\caption{A 95\% fit on the bounds expected for the branching
ratios of the decays $t\to cH$ and $t\to cZ$ if use is made of
the limits for the couplings $h_{r,l}$ and $g_{r,l}$ as given
in Fig.~\ref{figboundstchz}.
Two allowed regions are shown for $m_H=114$ and $145$ GeV:
inner and outer triangles respectively.}
\label{figboundstchzin}
\end{figure}

The possibility of extracting bounds on the effective vertices
tcV and tcH from loop observables has been studied in different
processes. Even though these bounds can not be considered
model-independent constraints, they may be regarded as order of
magnitude estimates which could be used in the search of new
physics effects in the following generation of colliders
\cite{reviews}.

The measurement of the inclusive branching ratio for the FCNC
process $b\to s\gamma$~\cite{alam} has been used to put
constraints on the $tc\gamma, tcg$ couplings~\cite{roberto,concha}.
These anomalous
couplings modify the coefficients of the operators $O_7$ and $O_8$
of the effective Hamiltonian for the $b\to s\gamma$ transition.
The known branching ratio for $t\to bW$ \cite{pdg} and the CLEO
bound on $b\to s\gamma$ place the limits $|\kappa_g|<0.9$ and
$|\kappa_\gamma|<0.16$, which can be translated into the bounds
$BR(t\to c\gamma)<2.2\times 10^{-3}$ and $BR(t\to cg)<3.4\times
10^{-2}$~\cite{roberto,concha}.
The $tcZ$ couplings $\kappa_r$ and
$\kappa_l$ given in Eq. (4) were bounded using several FCNC
low-energy processes such as $K_L\to \mu^+\mu^-$, $K_L-K_S$ mass
difference, $B^0-\bar{B}^0$ mixing and $B\to l^+l^-\gamma$, as
well as the oblique parameters $\rho$ and $S$: $\kappa_r<0.29$ and
$\kappa_l <0.05$ \cite{peccei}.
Since the transition amplitude for the
latter process is linear in  the $tcZ$ coupling, its decay rate
is very sensitive to the FCNC $tcZ$ coupling and produced a
stringent limit, $\kappa_l<0.05$~\cite{peccei}.

The FCNC couplings $tcZ$ and $tcH$ have also been constrained by
using the electroweak precision observables $\Gamma_Z$, $R_c$,
$R_b$, $R_l$, $A_c$ and the S/T oblique parameters~\cite{roberto05}.
The one-loop correction of these couplings to the decay modes $Z\to
c\bar c$ and $Z\to b\bar b$ are shown in Fig.~\ref{figdiagramszcc}.
Even though these vertices
enter in the Feynman diagrams 3(b)-(d) as a second order
perturbation, the known limits on the above precision
observables~\cite{pdg} impose significant constraints on
the $tcH$ and $tcZ$ couplings~\cite{roberto05}.

Figures~\ref{figboundstcz}~and~\ref{figboundstch}
show the $95\%$ C.L. limits on
the $g_l/g_r$ and $h_l/h_r$ FCNC top quark vertices and for
several values of the intermediate-mass Higgs boson. These limits
can be translated into the following bounds for the respective
branching ratios: $BR(t\to cZ)<1.6 \times 10^{-2}$ and $BR(t\to
cH)<0.9-29 \times 10^{-4}$ for
$116 \; GeV<mH<170 \; GeV$~\cite{roberto05}.

In particular, the limit on $Br(t\to cZ)$ is similar
to the bound recently reported  by the DELPHI Collaboration
\cite{abdalla}.
On the other hand, the contribution of the dimension-5
terms shown in Eq.(4) to the Feynman diagrams shown in
Fig.~\ref{figdiagramszcc} are
suppressed by a factor of order $(m_Z/m_t)^4$ with respect to the
corrections induced by the dimension-4 terms $tcZ$ and $tcH$.
This is the reason why the $BR(t\to c\gamma)$ does not acquire any
significant constraints from the above LEP precision
observables~\cite{roberto05}.
The interference of the $tcH$ and $tcZ$ contributions
to the rate $Z\to b\bar b$ does not improve the above limits
(Figs.~\ref{figboundstchz}~and~\ref{figboundstchzin}).

\section{Summary and outlook}

The FCNC decays of the top quark are very sensitive to physics
beyond the SM. Some extensions of the SM predict spectacular
enhancements on the FCNC branching ratios which are within the
reach of the LHC and the ILC. Searches for FCNC top-quark effects
in these colliders may thus constitute one of the best way to look
for physics beyond the SM: the experimental feasibility of
detecting such effects in top quark production and decays seems
to be better than the situation expected in the FCNC effects
of the $Z$ gauge boson~\cite{tavares}
and the Higgs boson~\cite{cruztoscano}
(or even in the case of the expected CP violating effects in
top quark physics~\cite{atwood}).

In Table \ref{ta1} we summarize the predictions that are
potentially visible at the LHC and ILC colliders for the
different models surveyed in the present review.
As we can appreciate, almost all
models include testable predictions for the FCNC channels which
will be accessible in these colliders. In this respect, even in
the optimistic situation of a positive detection of a FCNC decay
of the top quark, there would remain still the question to clear
up the nature of the virtual effects involved in the enhancement
of the respective FCNC decay. On the other hand,
if no new physics effect is ever found in this search,
an improvement on the experimental bounds
of FCNC top quark decays will provide a critical test of the
validity of the SM at the loop level.

\begin{table}
\tbl{Summary of the predictions that are potentially visible
at the LHC and ILC for BR-FCNC top-quark decay modes. References
to specific results are included in the text. The column for the
effective Lagrangian approach (ELA) includes the respective
bounds obtained for these decay modes from low-energy precision
measurements.}
{\begin{tabular}{@{}cccccccccc@{}} \toprule
Decay & THDM II & THDM III & MSSM &
$R\hspace{-0.18cm}\slash$ -MSSM &
TC2 & L-R & LR-SUSY & Extra q & ELA \\
\colrule
$t\to c\gamma$ & $10^{-7}$ &  $10^{-7}$ & $10^{-6}$ & $10^{-5}$ & 
 $10^{-7}$ &  & $10^{-6}$ & $10^{-8}$ & $10^{-3}$ \\ 
$t\to cZ$ &  $10^{-8}$ & $10^{-6}$ & $10^{-6}$ & $10^{-4}$ & 
$10^{-5}$ &  & $10^{-4}$ & $10^{-4}$ & $10^{-2}$ \\ 
$t\to cg$ & $10^{-5}$ & $10^{-4}$ & $10^{-4}$ & $10^{-3}$ & 
$10^{-5}$ &  & $10^{-5}$ & $10^{-7}$ & $10^{-2}$ \\ 
$t\to cH$ & $10^{-4}$ & $10^{-3}$ &  & $10^{-5}$ & 
 & $10^{-4}$ &  & $10^{-5}$ & $10^{-3}$ \\ 
$t\to cWW$ &  & $10^{-3}$ &  &  & 
$10^{-3}$ &  &  &  &  \\ 
$t\to cZZ$ &  & $10^{-3}$ &  &  & 
 &  &  &  &  \\ 
$t\to c\gamma \gamma $ & & $10^{-4}$ &  &  & 
 &  &  &  &  \\ 
$t\to c\gamma Z$ &  & $10^{-4}$ &  &  & 
 &  &  &  &  \\ 
$t\to cgg$ &  & $10^{-4}$ & $10^{-5}$ &  & 
 &  &  &  &  \\ 
$t\to cHg$ &  & $10^{-5}$ &  &  & 
 &  &  &  &  \\ 
\botrule
\end{tabular} \label{ta1}}
\end{table}

\newpage

\section*{Acknowledgments}

We thank Conacyt (M\'exico) and Colciencias (Colombia) for
support.

\end{document}